\documentclass[aps,prl,superscriptaddress,twocolumn,amsmath,amssymb,floatfix]{revtex4-1}
\usepackage{amssymb,amsfonts,amsmath,amstext,graphicx,hyperref,lipsum,empheq}
\usepackage[normalem]{ulem}
\hypersetup{
	colorlinks=true,
	linkcolor=blue,
	filecolor=magenta,      
	urlcolor=cyan,
}
\usepackage[dvipsnames]{xcolor}

\def\bra#1{\langle{#1}|}
\def\ket#1{|{#1}\rangle}
\def\braket#1{\langle{#1}\rangle}
\def\innerp#1#2{\langle{#1}\vert#2\rangle}

{\catcode`\|=\active
  \gdef\Braket#1{\begingroup
\mathcode`\|32768\let|\BraVert\left<{#1}\right>\endgroup}}
\def\BraVert{\egroup\,\mid\,\bgroup}

\definecolor{myblue}{rgb}{.8, .8, 1}

\begin{document}

\title{Near-Deterministic Weak-Value Metrology via Collective non-Linearity}

\author{Muthumanimaran Vetrivelan}
\email{muthu@phy.iitb.ac.in}
\affiliation{Department of Physics, Indian Institute of Technology-Bombay, Powai, Mumbai 400076, India}
\author{Sai Vinjanampathy}
\email{sai@phy.iitb.ac.in}
\affiliation{Department of Physics, Indian Institute of Technology-Bombay, Powai, Mumbai 400076, India}
\affiliation{Centre for Quantum Technologies, National University of Singapore, 3 Science Drive 2, 117543 Singapore, Singapore}

\date{\today}

\begin{abstract}
Weak-value amplification employs postselection to enhance the measurement of small parameters of interest. The amplification comes at the expense of reduced success probability, hindering the utility of this technique as a tool for practical metrology. Following other quantum technologies that display a quantum advantage, we formalize a quantum advantage in the success probability and present a scheme based on non-linear collective Hamiltonians that shows a super-extensive growth in success probability while simultaneously displaying an extensive growth in the weak value. We propose an experimental implementation of our scheme. 
\end{abstract} 
\maketitle

\makeatletter

\paragraph{Introduction.---} 
Quantum metrology employs non-classical correlations and Hamiltonian interactions to enhance parameter estimation \cite{Giovannetti2011,paula}. An example of a metrology scheme with demonstrable advantages to measure weak signals in the presence of technical noise sources \cite{PhysRevX.4.011031} and also for saturated signals \cite{Animesh} is weak-value metrology. Weak-value amplification (WVA) uses postselection to enhance measurements of small parameters via amplification \cite{PhysRevLett.60.1351,PhysRevLett.66.1107,Kofman2012,Dressel2014,kedem_wv}. Such postselection involves rejecting measurement outcomes and there exists a trade-off between the postselection probability of the desired measurement and the amplification produced by the apparatus. This trade-off has always limited the applicability of weak-value metrology. Given its practical applications \cite{ben_dixon,thermometry,sibasish,ghosh2019weak}, it is desirable to have (at least) linear scaling in both quantities, since this promises that a modest increase in the number of atoms participating in the collective interaction will amplify the signal while simultaneously making the desired measurements more probable. Such a solution was lacking until now. In this manuscript, we show that non-linear collective Hamiltonians can enhance existing WVA schemes substantially. We demonstrate a linear enhancement in the weak-value $\mathcal{A}_w$ and a simultaneous quadratic enhancement in the success probability $P_s$, both defined below. Furthermore, we discuss a near-deterministic WVA scheme that shows a quadratic enhancement of $\mathcal{A}_w$. Finally,  we propose an experimental scheme that realizes the initial state preparation, the final system state measurement and an implementation of the non-linear WVA Hamiltonian to realize this advantage using existing experimental resources.

\paragraph{Success Probability and Weak Value.---} Consider the bipartite Hamiltonian
$H_I=g A \otimes B\delta(t-t')$, where $A$($B$) is a Hermitian operator defined in the system(meter) Hilbert space and $g$ is the extremely small parameter of interest. The initial state $\ket{\Psi_{i}}=\ket{\psi_{i}}\otimes\ket{\phi_{i}}$ is allowed to weakly interact for a small time and is then postselected in a final system state $\ket{\psi_{f}}$. The resulting postselected or ``kicked" meter state is given by 
$\ket{\phi_{f}}=M\ket{\phi_{i}}/\vert\vert M\ket{\phi_{i}}\vert\vert$, where $M=\exp(-ig\mathcal{A}_{w}B)$ is the non-unitary evolution defined in terms of the weak value $\mathcal{A}_{w}$ \cite{PhysRevLett.66.1107}, given by
\begin{equation}
\mathcal{A}_{w}=\frac{\bra{\psi_{f}}A\ket{\psi_{i}}}{\innerp{\psi_{f}}{\psi_{i}}} \label{wv1}.
\end{equation}

We will refer to the choice of the initial state of the system, the final state of the system and the initial state of the meter as the \textit{weak value strategy}. Note that the $\mathcal{A}_w$ is inversely proportional to the square-root of the success probability $P_s=\vert\innerp{\psi_{f}}{\psi_{i}}\vert^2$. A small success probability hence implies a large amplification. WVA hence allows for the small value $g$ to be amplified due to the effect of the measurement of the system. In general $\mathcal{A}_{w}$ can be a complex number and greater than the expectation value of the observable $A$.  For unbiased estimators, when the weak value is chosen to be purely imaginary  \cite{pang2014entanglement} the expectation value of a meter observable $R$ is approximately given by  $\braket{R}_{\ket{\phi_{f}}} \approx 2g[Im(\mathcal{A}_{w})Re(\alpha)]$ (where $\alpha = \braket{RB}_{\ket{\phi_{i}}}$, offering the potential amplification of $g$ in the measured signal \cite{pang2014entanglement}. For example, if $A=\sigma_x$, $\ket{\psi_i}=\ket{\downarrow}$ and $\ket{\psi_f}=\ket{\theta}=\sin(\theta)\ket{\downarrow}+\exp(i\frac{\pi}{2})\cos(\theta)\ket{\uparrow}$, then $\mathcal{A}_w\approx i/\theta$ for small $\theta$. This implies that the choice of the postselected state to be nearly orthogonal to the initial state amplifies $g$ to a larger number which can then be read from the kicked meter state. The central trade-off of the not deterministic WVA technique is that enhancing weak value comes at a corresponding loss of the success probability of the postselection event. This hinders the utility of the WVA procedure in practice, since the successful detector clicks are always competing with the amplification. One may seek to remedy this by having multiple probes.

If measurements are performed on $2j$ independent copies, the success probability of at least one positive click scales extensively in the number of independent ancilla \cite{PhysRevA.77.032101} since
\begin{equation}
P_{s}^{(2j)}=1-(1-P_{s})^{2j} \approx 2jP_{s},  \label{pab0}
\end{equation}
where we denote the success probability associated with multiple probes as $P_{s}^{(2j)}$. This presupposes that at the end of these $2j$ detector clicks, we are satisfied with just one of the experimental runs returning an amplification. If we require that all $2j$ independent measurements simultaneously produce weak value amplification, then the corresponding success probability decreases exponentially.  A natural question that arises is whether this extensive scaling of $P_s$ can be improved while enhancing the weak value. Since entanglement provides an advantage in computing and allows us to go beyond shot-noise variance in metrology, it is reasonable to expect an entanglement advantage in WVA as well. Furthermore, since an advantage due to collective interactions  has been observed in disparate technologies such as metrology \cite{Caves,zwierz2010general} and energy storage \cite{binder2015quantacell,campaioli2017enhancing,rossini2019many,PhysRevLett.122.047702}, it is reasonable to expect collective interactions to lead to an advantage in weak-value metrology. This was explored by Pang and Brun \cite{pang2014entanglement} whose proposal leads to sub-linear scaling in $\mathcal{A}_w$ with a simultaneous linear enhancement in $P_s$, which we briefly review. 
\begin{figure}[htp!]
  \includegraphics[width=0.9\linewidth]{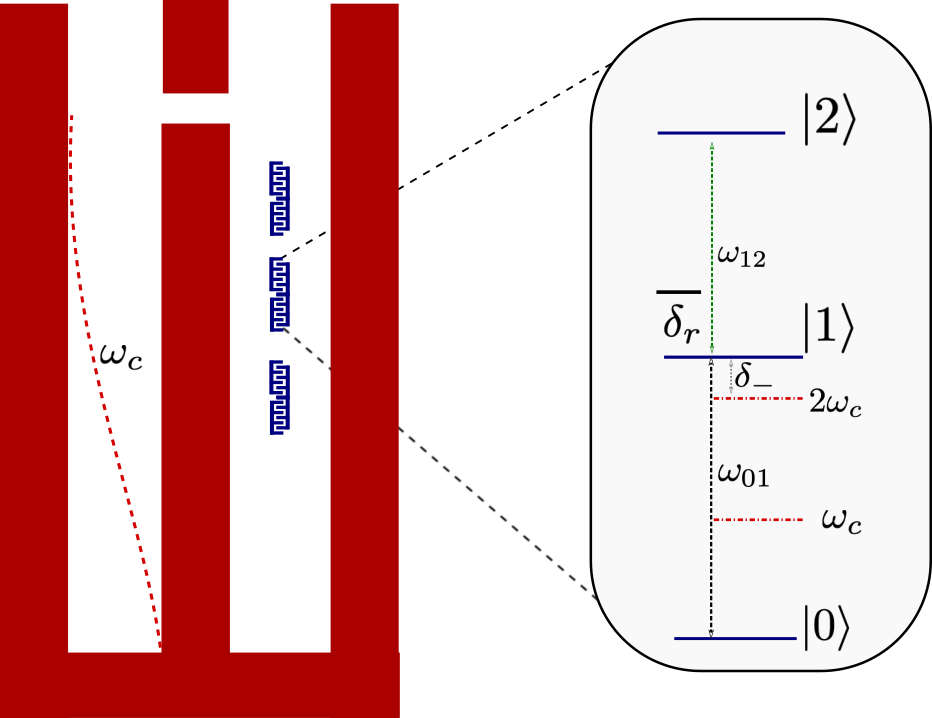}
  \caption{Circuit QED analogue of our model with the energy levels of the atom (transmon) and the first two levels of the resonator modes denoted. The resonator mode at $\omega_c$ is two-photon detuned from $\omega_{01}$ as shown. Interaction with another far-detuned readout mode is indicated at the detuning $\delta_r$.}
  \label{energy_levels}
\end{figure}

\paragraph{Optimized Trade-off for Linear Hamiltonians.---} The postselection probability, the weak value and the variance of the Hamiltonian in the initial state are related to each other \cite{pang2014entanglement}. Hence an optimization of $P_s^{(2j)}$ or $\mathcal{A}_w$ can be performed separately, each time at the expense of the other. To optimize $P_s^{(2j)}$, Eq.~\eqref{wv1} can be modified to read
$\bra{\psi_{f}}(A-\mathcal{A}_{w})\ket{\psi_{i}}=0$. 
The postselected state is hence orthogonal to the state $(A-\mathcal{A}_{w})\ket{\psi_{i}}$.  The optimal postselected state can be found using Gram-Schmidt orthogonalization and the corresponding maximum $P_s^{(2j)}$ is
\begin{equation}
\max P_{s}^{(2j)} \approx \frac{Var(A)_{\ket{\psi_{i}}}}{\vert\mathcal{A}_{w}\vert^2} \label{pab3},
\end{equation}

where $Var(A)_{\ket{\psi_{i}}}=\bra{\psi_{i}}A^2\ket{\psi_{i}}-[\bra{\psi_{i}}A\ket{\psi_{i}}]^2$. This equation is valid for $\vert\mathcal{A}_{w}\vert\gg \max\langle A\rangle$ .  For the collective operator $A=J_z$, maximizing the variance over an arbitrary initial state $\ket{\psi_{i}}$ yields \cite{Giovannetti1330}
$\max  Var(J_z)_{\ket{\psi_{i}}} =j^2$, corresponding to the state
\begin{equation}
\ket{\psi_{i}}\propto \ket{j,j}+\ket{j,-j}, \label{pab5}
\end{equation}

where normalization constant has been omitted.  This initial state is well known for its advantage in various quantum metrology protocols \cite{Giovannetti2011}. For a fixed value of $\mathcal{A}_{w}$, the postselection probability hence scales as $j^2$. This is $j$ times better than using uncoupled ancillas which shows linear scaling in $j$.  Using the orthogonalisation argument presented above, the final postselection state for linear collective Hamiltonians takes of the form
 \begin{equation}
\ket{\psi_{f}}\propto (j+\mathcal{A}_{w})\ket{j,j}+(j-\mathcal{A}_{w})\ket{j,-j}.     
\label{pss}
 \end{equation}
 Hence the entanglement of the initial/final state and the collective Hamiltonian interactions are leveraged to produce a quantifiable advantage in the success probability. To compare collective and individual strategies for weak value amplification, we define the  \textit{success probability advantage} of collective strategies as
\begin{equation}
\sigma:=\frac{P_s^{(2j)}}{2jP_s}.
\end{equation} 
While an uncorrelated weak-value strategy yields an advantage $\sigma\propto 1$, the aforementioned entanglement strategy offers $\sigma\propto j$ for a fixed $\mathcal{A}_w$. Clearly the limitation of this strategy is the extensive scaling of $\sigma$ comes at the cost of holding $\mathcal{A}_w$ fixed, though a large number of particles are participating in the dynamics. 

On the other hand, WVA can be optimized by holding $\sigma$ fixed. Consider a postselection state $\ket{\psi_{f}}$ of the form

\begin{equation}
\ket{\psi_{f}}=\sqrt{P_{s}^{(2j)}}\ket{\psi_{i}}+\sqrt{1-P_{s}^{(2j)}}\ket{\psi_{i}^{\perp}}, \label{pab6}
\end{equation}

where $\ket{\psi_{i}^{\perp}}$ is a state that is orthogonal to the initial state 
$\ket{\psi_{i}}$.  The weak value in Eq.~\eqref{wv1} becomes

\begin{equation}
\mathcal{A}_{w}=\bra{\psi_{i}}A\ket{\psi_{i}}+\sqrt{\frac{1-P_{s}^{(2j)}}{P_{s}^{(2j)}}}\bra{\psi_{i}^{\perp}}A\ket{\psi_{i}}.  \label{pab7}
\end{equation}
For small $P_s$, it can be shown that 
\begin{equation}
\max \vert\mathcal{A}_{w}\vert \approx \sqrt{\frac{Var(A)_{\ket{\psi_{i}}}}{P_{s}^{(2j)}}}. \label{pab8}
\end{equation}

As noted in Eq.~\eqref{pab3}, the maximum weak value depends only on the variance of observable $A$ with respect to initial state.  By fixing $\sigma\propto 1$, the maximum attainable weak-value scaling is $\mathcal{A}_w\propto \sqrt{j}$.

To summarize the case of linear collective Hamiltonians, there is a definite trade off between either maximizing post-measurement quantum advantage $\sigma$ or gaining a collective advantage in the weak value $\mathcal{A}_w$. Different strategies involving linear Hamiltonians can either produce an extensive success probability advantage for fixed $\mathcal{A}_w$ or no advantage for the case where $\mathcal{A}_w\propto \sqrt{j}$. We now show that non-linear Hamiltonian metrology offers a simultaneous collective advantage $\sigma$ and an enhancement in $\mathcal{A}_w$ that scales favorably with $j$. 

\begin{figure}
  \includegraphics[width=0.95\linewidth]{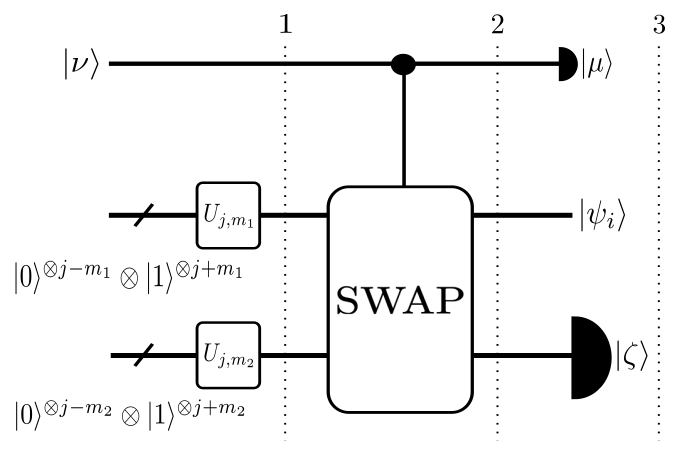}
  \caption{Deterministic Dicke state preparation involves the action of the unitary operators $U_{j,m_1}$ and $U_{j,m_2}$ on the computational state prepare Dicke states.  The state after passing through the control-SWAP gate is projected onto $\ket{\mu}\bra{\mu}\otimes I \otimes \ket{\zeta}\bra{\zeta}$. The non-zero overlaps $\braket{j,m_1\vert\zeta}$ and $\braket{j,m_2\vert\zeta}$ are used to define the state $\ket{\mu}$ which produces the desired initial state as shown. } 
  \label{input}
\end{figure}

\paragraph{Non-Linearity and Joint Optimization.---} Consider the Hamiltonian
\begin{equation}
H=g({J}^2-{J}_{z}^2) \otimes {a}^{\dagger}{a}, \label{oc1}
\end{equation}
whose physical implementation is discussed below. The maximum and minimum eigenvalues of the system Hamiltonian are $j(j+1)$ and $j$ respectively \cite{bartschi2019deterministic} correspond to the eigenstates $\ket{j,0}$ and $\ket{j, \pm j}$. Choosing $\ket{j,-j}$ as the lower eigenstate, the initial state is taken as the equal superposition of these eigenstates, namely
\begin{equation}
\ket{\psi_{i}}\propto \ket{j,0}+\ket{j,-j}. \label{oc11}
\end{equation} 
Following the analysis outlined for linear Hamiltonians, the maximum variance with $A=J^2-J_z^2$ is
\begin{equation}
\max Var(A)_{\ket{\psi_{i}}} = \frac{j^4}{4}. \label{oc3}
\end{equation} 
Since the maximum value of $\mathcal{A}_{w}$ in Eq.~\eqref{pab8} depends on the square root of variance over the postselection probability, if we fix $P_{s}^{(2j)} = \kappa j^2\ll 1$, the maximum weak value will scale as $\mathcal{A}_{w} \propto j$. This means that simultaneously we have achieved superextensive scaling in the success probability denoted by $\sigma \propto j$ and extensive weak value amplification due to collective non-linearity. This is one of the main results of our manuscript. The corresponding postselected state (see Appendix A) can be evaluated to be
\begin{equation}
\ket{\psi_{f}} \propto (\sqrt{\kappa}j+1)\ket{j,0} + (\sqrt{\kappa}j-1)\ket{j,-j}.  \label{oc5}
\end{equation}

\paragraph{Near-Deterministic Strategy.---} On the other hand, we can optimize the final state differently to obtain a near-deterministic weak-value metrology protocol. Let us consider Eq.~\eqref{pab7}. once again, but seek a near-deterministic protocol by setting $P_{s}^{(2j)}=1-\epsilon$ . If we choose $\ket{\psi_{f}}\propto (1+\sqrt{\epsilon})\ket{j,0}+(1-\sqrt{\epsilon})\ket{j,-j}$, with the same initial state as before, we obtain a nearly deterministic protocol with quadratic scaling in $\mathcal{A}_{w}$ simultaneously.
This quantum advantage is predicated on (a) an implementation of the non-linear Hamiltonian in Eq.~\eqref{oc1} and  (b) preparation of the initial state and measurement in the final system state. We now detail an experimental proposal to implement this. 

\paragraph{Implementation of non-Linear Hamiltonian.---} 
Consider the two-photon Tavis-Cummings Hamiltonian namely
\begin{equation}
    H_0=g_0(J_{+}a^2e^{i\delta_{-}t}+J_{-}a^{\dag^2}e^{-i\delta_{-}t}).
    \label{2p-tc}
\end{equation}
Such a two-photon collective Hamiltonian can be implemented by generalizing existing two-photon Jaynes-Cummings \cite{Solano_two_photon_dicke_model,Garbe_2020} models to include several atoms or modifying an existing Dicke model implementation to induce a two-photon term \cite{fink2009dressed,mlynek2014observation}. We propose the second method, wherein $2j$ transmon atoms \cite{Song574} are coupled to a resonator mode whose fundamental frequency is at $\omega_c$. Since the cavity frequency is close to half the transmon's frequency $\omega_{01}$, one needs to ensure that any higher modes of the cavity are sufficiently detuned from $\omega_{01}$. This can be achieved by either using lumped-element resonators \cite{vijay2011observation} which effectively only have one mode, or by choosing a quarter-wave transmission line resonator \cite{pozar1990microwave}  whose second harmonic is at three times the fundamental frequency. This is depicted in Fig.~(\ref{energy_levels}) alongside the energy level of a (different) cavity used to perform the qubit transformations highlighted in the next section.

In order to derive the Hamiltonian from a two-photon Rabi model \cite{Solano_two_photon_rabi_model}, the frequency $\omega_{01}$ and the cavity frequency $\omega_c$ are related to the bare coupling via $g_0<  \vert\delta_{-}\vert \ll \delta_+$. Here $\delta_{\mp}=\omega_0\mp 2\omega_c$ is the two-photon detuning (sum frequency) and we have already dropped the counter-rotating terms owing to the magnitude of $\delta_+$ to arrive at Eq.~(\ref{2p-tc}). In the far-detuned regime $g_0\ll \delta_{-}$, we can write the effective Hamiltonian \cite{james2007effective,PhysRevA.95.032124} as Eq.(\ref{oc1}) with the identification $g\equiv-4g_0^2/\delta_{-}$. This Hamiltonian can be understood to induce a (collective) state dependent Bloch-Siegert \cite{forn2010observation} like shift in the harmonic spectrum. Such a Hamiltonian represents a non-linear interaction between atoms mediated by a cavity and is crucially different from the Tavis-Cummings model where twisting Hamiltonians \cite{ma2011quantum,paula} are produced by adiabatically eliminating the cavity. We note that the coupling in  Eq.~(\ref{2p-tc}) is significantly small and hence the spectral collapse \cite{Garbe_2020} of the Hamiltonian in ultra-strong coupling regime is avoided. A small magnetic field would displace the energy levels represented by $\omega_{01}$ linearly, which can be inferred from a corresponding weak-value amplified shift in the measured value of $4g_0^2/\delta_-$.

\begin{figure}
  \includegraphics[width=0.95\linewidth]{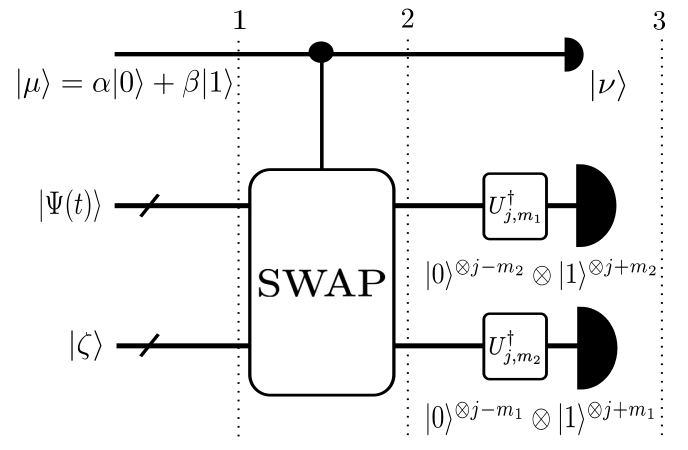}
  \caption{The time-evolved system-meter state $\ket{\Psi(t)}$ enters the circuit where $\ket{\mu}$ and $\ket{\zeta}$ are ancilla and reference states defined in the text. The final state is projected onto the joint quantum state $\ket{\nu}\otimes\ket{j,m_1}\otimes\ket{j,m_2}$ which results in the postselection of the system in the state $\ket{\psi_f}$.}
  \label{measurement}
\end{figure}

\paragraph{State Preparation \& Measurement.---} Now we propose a scheme to prepare the initial state and perform the measurement in the final state. The solution is presented as multi-qubit circuit implementations, which may be implemented by using an additional resonator mode. The initial state for all of the weak value strategies considered is the superposition of two Dicke states, which we can write as $\alpha\ket{j,m_1}+\beta\ket{j,m_2}$. Dicke states can be prepared with an efficient deterministic quantum algorithm \cite{bartschi2019deterministic}.  We combine this with a probabilistic protocol for implementing superposition of two pure states \cite{PhysRevLett.116.110403,exptimplinitial} if the overlap with a reference state $\ket{\zeta}$ is known. The choice of the reference state $\ket{\zeta}$ is arbitrary and constrained only to have non-zero overlap with the two states being superposed. The two states $\ket{j,m_{1,2}}$ are input into the device shown in Fig.~(\ref{input}). An ancilla is initialized in the state  $\alpha\ket{0}+\beta\ket{1}$ and the control-SWAP gate is implemented. Following this, the second output is projected onto the state $\ket{\zeta}$ and the ancilla qubit is projected onto a state whose coefficients are drawn from $\ket{\zeta}$ \cite{PhysRevLett.116.110403} (see Appendix B). Since a projector onto the state $\ket{\zeta}$ has to be implemented, a judicious choice of the state incorporates ease of implementation.  To prepare the initial state in Eq.~\eqref{pab5} we choose $\ket{\zeta}=\ket{+}^{\otimes 2j}$. Although the success probability of this state preparation is $2^{-4j} \binom{2j}{j}^2$, this procedure can serve as an offline resource for the initial state. Once prepared, the success probability of the metrology scheme follows the scaling arguments we outline.

Finally, we address the task of measurement in the states $\ket{\zeta}$ and $\ket{\psi_f}$. The circuit presented in Fig.~(\ref{measurement}) implements the projector $\ket{\psi_f}\bra{\psi_f}$ on the time-evolved state (see Appendix C for a detailed derivation). This circuit is the standard modification of the superposition circuit that exchanges sources with detectors. Our modified scheme hence starts with the initial state $\ket{R_1}=\ket{\mu}\otimes\ket{\Psi(t)}\otimes\ket{\zeta}$, where $\ket{\Psi(t)}$ is the time evolved system-meter state. We then pass it through the circuit shown in Fig.~(\ref{measurement}). The evolved state $\ket{R_2}$ is projected onto the system basis $\ket{\nu}\otimes\ket{j,m_1}\otimes\ket{j,m_2}$. The successful projection onto $\ket{\psi_f}$ happens with probability $\tilde{P}_{s}\propto [\alpha^{*}\braket{\Psi(t)\vert j,m_1}+\beta^{*}\braket{\Psi(t)\vert j,m_2}][\alpha\braket{j,m_1\vert\Psi(t)}+\beta\braket{j,m_2\vert\Psi(t)}]$. For the state in  Eq.~\eqref{oc5}, we choose $m_1=0$ and $m_2=-j$ which yields the same scaling  $\tilde{P}_{s}\propto j^2$ as $P^{(2j)}_s$.

\paragraph{Fisher Analysis.---} postselecting on a rare outcome always involves discarding many measurement results. The question that arises is to establish under what conditions the postselected states have \textit{a majority} of the information available from the full measurement of the joint pre-measurement state. The information content is quantified by the quantum Fisher information $I(g)$ (QFI) associated with the joint state is  \cite{PhysRevA.91.062107,PhysRevLett.72.3439} given by
\begin{equation}
I(g)=4\frac{d\bra{\Psi_{g}}}{dg}\frac{d\ket{\Psi_{g}}}{dg}-4\bigg|{\frac{d\bra{\Psi_{g}}}{dg}\ket{\Psi_{g}}}\bigg|^2, \label{pab9}
\end{equation}

where $\ket{\Psi_{g}}=\exp(-igH)\ket{\Psi_{i}}$. For an unbiased estimator, it is well known \cite{PhysRevLett.72.3439} that the variance of the unknown parameter scales inversely with the QFI namely $Var(g)\propto I^{-1}(g)$. For the bipartite Hamiltonian $g A\otimes B$ the QFI is evaluated as $I(g) = 4[\braket{A^2}\braket{B^2}-(\braket{A}\braket{B})^2]$. For the initial state of the system given by Eq.~\eqref{oc11} and meter being in the coherent state $\ket{\phi_{i}}=\ket{\eta}$, this is evaluated to be

\begin{eqnarray}
I(g) & = & 4\bigg[\frac{1}{2}(j^4+2j^3+2j^2)(\vert\eta\vert^4+\vert\eta\vert^2) \nonumber \\
& &-\frac{1}{4}(j^4+4j^3+4j^2)\vert\eta\vert^4 \bigg], \nonumber\\
& \approx & 2j^4\vert\eta\vert^2,
\end{eqnarray}
for $\vert\eta\vert\ll 1$.

 Furthermore, we also investigate if the QFI present in the measurement of the state in Eq.~\eqref{oc5} saturates the total QFI available to the initial state.  Let $I^{'}(g)$ denote the quantum Fisher information of the kicked meter state after postselection of the system state.  The ratio of the post-measurement Fisher information to the total QFI in this case is evaluated to be 

\begin{equation}
\frac{I^{'}(g)}{I(g)}\approx \frac{1}{2}(1-\vert\eta g\mathcal{A}_{w}\vert^2), \label{oc7}
\end{equation}

where $\vert\eta g\mathcal{A}_{w}\vert \ll 1$.  Hence though only half of all of the information is available from the rarely postselected states from the full measurement, this is a constant reduction in the total QFI independent of the number of particles. A Fisher analysis of the near-deterministic scheme provides a similar scaling.

\paragraph{Conclusions.---} One of the drawbacks of WVA is that detector clicks are competing with the desired amplification. In this work, we describe a metrology scheme that simultaneously enhances the success probability and weak value. Furthermore, we demonstrate a collective advantage from non-linear Hamiltonians and prove that the success probability advantage scales linearly $\sigma\propto j$. This extends the well-known advantage of non-linear Hamiltonians \cite{Caves} to WVA metrology. This super-extensive enhancement of the success probability implies that with a modest enhancement in the number of spins $2j$, there can be a dramatic enhancement of the success probability and the amplified weak value. We propose experimental implementations of the crucial steps of the protocol, namely the preparation of the initial states, the measurement in the final state and the implementation of the proposed interaction Hamiltonian. The initial and reference states are prepared as an offline resource which does not impact the success probability of the metrology scheme. Our scheme makes it possible to perform weak-value metrology with linear enhancement in weak-value and quadratic enhancement in success probability. Such an enhancement can be further generalized to other higher order non-linear Hamiltonians using our theoretical approach and the corresponding experimental proposal. An information theoretic analysis of both our schemes show that approximately half of the available Fisher information is collected by our scheme, a ratio that does not scale with the number of atoms. Finally, we formalize the quantum advantage due to collective interactions and non-linear Hamiltonians via the success probability advantage $\sigma$.

This simultaneous enhancement can be used for practical metrology protocols which still retain the known advantages of WVA metrology, in the amplifying exceedingly small parameters, in the presence of technical noise sources and in the presence of detector saturation while boosting the underlying probabilities. Such enhancement will find applications in sensing various physical parameters of interest.

\paragraph{Acknowledgements.---} S.V. acknowledges support from the DST-SERB Early Career Research Award (ECR/2018/000957) and DST-QUEST grant number DST/ICPS/QuST/Theme-4/2019. S.V. thanks Yaron Kedem, Prasanna B. Venkatesh and R.Vijay for insightful discussions. 


	\bibliographystyle{apsrev4-1}

\begin{thebibliography}{39}%
\makeatletter
\providecommand \@ifxundefined [1]{%
 \@ifx{#1\undefined}
}%
\providecommand \@ifnum [1]{%
 \ifnum #1\expandafter \@firstoftwo
 \else \expandafter \@secondoftwo
 \fi
}%
\providecommand \@ifx [1]{%
 \ifx #1\expandafter \@firstoftwo
 \else \expandafter \@secondoftwo
 \fi
}%
\providecommand \natexlab [1]{#1}%
\providecommand \enquote  [1]{``#1''}%
\providecommand \bibnamefont  [1]{#1}%
\providecommand \bibfnamefont [1]{#1}%
\providecommand \citenamefont [1]{#1}%
\providecommand \href@noop [0]{\@secondoftwo}%
\providecommand \href [0]{\begingroup \@sanitize@url \@href}%
\providecommand \@href[1]{\@@startlink{#1}\@@href}%
\providecommand \@@href[1]{\endgroup#1\@@endlink}%
\providecommand \@sanitize@url [0]{\catcode `\\12\catcode `\$12\catcode
  `\&12\catcode `\#12\catcode `\^12\catcode `\_12\catcode `\%12\relax}%
\providecommand \@@startlink[1]{}%
\providecommand \@@endlink[0]{}%
\providecommand \url  [0]{\begingroup\@sanitize@url \@url }%
\providecommand \@url [1]{\endgroup\@href {#1}{\urlprefix }}%
\providecommand \urlprefix  [0]{URL }%
\providecommand \Eprint [0]{\href }%
\providecommand \doibase [0]{http://dx.doi.org/}%
\providecommand \selectlanguage [0]{\@gobble}%
\providecommand \bibinfo  [0]{\@secondoftwo}%
\providecommand \bibfield  [0]{\@secondoftwo}%
\providecommand \translation [1]{[#1]}%
\providecommand \BibitemOpen [0]{}%
\providecommand \bibitemStop [0]{}%
\providecommand \bibitemNoStop [0]{.\EOS\space}%
\providecommand \EOS [0]{\spacefactor3000\relax}%
\providecommand \BibitemShut  [1]{\csname bibitem#1\endcsname}%
\let\auto@bib@innerbib\@empty
\bibitem [{\citenamefont {Giovannetti}\ \emph {et~al.}(2011)\citenamefont
  {Giovannetti}, \citenamefont {Lloyd},\ and\ \citenamefont
  {Maccone}}]{Giovannetti2011}%
  \BibitemOpen
  \bibfield  {author} {\bibinfo {author} {\bibfnamefont {V.}~\bibnamefont
  {Giovannetti}}, \bibinfo {author} {\bibfnamefont {S.}~\bibnamefont {Lloyd}},
  \ and\ \bibinfo {author} {\bibfnamefont {L.}~\bibnamefont {Maccone}},\ }\href
  {\doibase 10.1038/nphoton.2011.35} {\bibfield  {journal} {\bibinfo  {journal}
  {Nature Photonics}\ }\textbf {\bibinfo {volume} {5}},\ \bibinfo {pages} {222}
  (\bibinfo {year} {2011})}\BibitemShut {NoStop}%
\bibitem [{\citenamefont {Degen}\ \emph {et~al.}(2017)\citenamefont {Degen},
  \citenamefont {Reinhard},\ and\ \citenamefont {Cappellaro}}]{paula}%
  \BibitemOpen
  \bibfield  {author} {\bibinfo {author} {\bibfnamefont {C.~L.}\ \bibnamefont
  {Degen}}, \bibinfo {author} {\bibfnamefont {F.}~\bibnamefont {Reinhard}}, \
  and\ \bibinfo {author} {\bibfnamefont {P.}~\bibnamefont {Cappellaro}},\
  }\href {\doibase 10.1103/RevModPhys.89.035002} {\bibfield  {journal}
  {\bibinfo  {journal} {Rev. Mod. Phys.}\ }\textbf {\bibinfo {volume} {89}},\
  \bibinfo {pages} {035002} (\bibinfo {year} {2017})}\BibitemShut {NoStop}%
\bibitem [{\citenamefont {Jordan}\ \emph {et~al.}(2014)\citenamefont {Jordan},
  \citenamefont {Mart\'{\i}nez-Rinc\'on},\ and\ \citenamefont
  {Howell}}]{PhysRevX.4.011031}%
  \BibitemOpen
  \bibfield  {author} {\bibinfo {author} {\bibfnamefont {A.~N.}\ \bibnamefont
  {Jordan}}, \bibinfo {author} {\bibfnamefont {J.}~\bibnamefont
  {Mart\'{\i}nez-Rinc\'on}}, \ and\ \bibinfo {author} {\bibfnamefont {J.~C.}\
  \bibnamefont {Howell}},\ }\href {\doibase 10.1103/PhysRevX.4.011031}
  {\bibfield  {journal} {\bibinfo  {journal} {Phys. Rev. X}\ }\textbf {\bibinfo
  {volume} {4}},\ \bibinfo {pages} {011031} (\bibinfo {year}
  {2014})}\BibitemShut {NoStop}%
\bibitem [{\citenamefont {Xu}\ \emph {et~al.}(2020)\citenamefont {Xu},
  \citenamefont {Liu}, \citenamefont {Datta}, \citenamefont {Knee},
  \citenamefont {Lundeen}, \citenamefont {Lu},\ and\ \citenamefont
  {Zhang}}]{Animesh}%
  \BibitemOpen
  \bibfield  {author} {\bibinfo {author} {\bibfnamefont {L.}~\bibnamefont
  {Xu}}, \bibinfo {author} {\bibfnamefont {Z.}~\bibnamefont {Liu}}, \bibinfo
  {author} {\bibfnamefont {A.}~\bibnamefont {Datta}}, \bibinfo {author}
  {\bibfnamefont {G.~C.}\ \bibnamefont {Knee}}, \bibinfo {author}
  {\bibfnamefont {J.~S.}\ \bibnamefont {Lundeen}}, \bibinfo {author}
  {\bibfnamefont {Y.-q.}\ \bibnamefont {Lu}}, \ and\ \bibinfo {author}
  {\bibfnamefont {L.}~\bibnamefont {Zhang}},\ }\href {\doibase
  10.1103/PhysRevLett.125.080501} {\bibfield  {journal} {\bibinfo  {journal}
  {Phys. Rev. Lett.}\ }\textbf {\bibinfo {volume} {125}},\ \bibinfo {pages}
  {080501} (\bibinfo {year} {2020})}\BibitemShut {NoStop}%
\bibitem [{\citenamefont {Aharonov}\ \emph {et~al.}(1988)\citenamefont
  {Aharonov}, \citenamefont {Albert},\ and\ \citenamefont
  {Vaidman}}]{PhysRevLett.60.1351}%
  \BibitemOpen
  \bibfield  {author} {\bibinfo {author} {\bibfnamefont {Y.}~\bibnamefont
  {Aharonov}}, \bibinfo {author} {\bibfnamefont {D.~Z.}\ \bibnamefont
  {Albert}}, \ and\ \bibinfo {author} {\bibfnamefont {L.}~\bibnamefont
  {Vaidman}},\ }\href {\doibase 10.1103/PhysRevLett.60.1351} {\bibfield
  {journal} {\bibinfo  {journal} {Phys. Rev. Lett.}\ }\textbf {\bibinfo
  {volume} {60}},\ \bibinfo {pages} {1351} (\bibinfo {year}
  {1988})}\BibitemShut {NoStop}%
\bibitem [{\citenamefont {Ritchie}\ \emph {et~al.}(1991)\citenamefont
  {Ritchie}, \citenamefont {Story},\ and\ \citenamefont
  {Hulet}}]{PhysRevLett.66.1107}%
  \BibitemOpen
  \bibfield  {author} {\bibinfo {author} {\bibfnamefont {N.~W.~M.}\
  \bibnamefont {Ritchie}}, \bibinfo {author} {\bibfnamefont {J.~G.}\
  \bibnamefont {Story}}, \ and\ \bibinfo {author} {\bibfnamefont {R.~G.}\
  \bibnamefont {Hulet}},\ }\href {\doibase 10.1103/PhysRevLett.66.1107}
  {\bibfield  {journal} {\bibinfo  {journal} {Phys. Rev. Lett.}\ }\textbf
  {\bibinfo {volume} {66}},\ \bibinfo {pages} {1107} (\bibinfo {year}
  {1991})}\BibitemShut {NoStop}%
\bibitem [{\citenamefont {Kofman}\ \emph {et~al.}(2012)\citenamefont {Kofman},
  \citenamefont {Ashhab},\ and\ \citenamefont {Nori}}]{Kofman2012}%
  \BibitemOpen
  \bibfield  {author} {\bibinfo {author} {\bibfnamefont {A.~G.}\ \bibnamefont
  {Kofman}}, \bibinfo {author} {\bibfnamefont {S.}~\bibnamefont {Ashhab}}, \
  and\ \bibinfo {author} {\bibfnamefont {F.}~\bibnamefont {Nori}},\ }\href
  {\doibase https://doi.org/10.1016/j.physrep.2012.07.001} {\bibfield
  {journal} {\bibinfo  {journal} {Physics Reports}\ }\textbf {\bibinfo {volume}
  {520}},\ \bibinfo {pages} {43 } (\bibinfo {year} {2012})},\ \bibinfo {note}
  {nonperturbative theory of weak pre- and post-selected
  measurements}\BibitemShut {NoStop}%
\bibitem [{\citenamefont {Dressel}\ \emph {et~al.}(2014)\citenamefont
  {Dressel}, \citenamefont {Malik}, \citenamefont {Miatto}, \citenamefont
  {Jordan},\ and\ \citenamefont {Boyd}}]{Dressel2014}%
  \BibitemOpen
  \bibfield  {author} {\bibinfo {author} {\bibfnamefont {J.}~\bibnamefont
  {Dressel}}, \bibinfo {author} {\bibfnamefont {M.}~\bibnamefont {Malik}},
  \bibinfo {author} {\bibfnamefont {F.~M.}\ \bibnamefont {Miatto}}, \bibinfo
  {author} {\bibfnamefont {A.~N.}\ \bibnamefont {Jordan}}, \ and\ \bibinfo
  {author} {\bibfnamefont {R.~W.}\ \bibnamefont {Boyd}},\ }\href {\doibase
  10.1103/RevModPhys.86.307} {\bibfield  {journal} {\bibinfo  {journal} {Rev.
  Mod. Phys.}\ }\textbf {\bibinfo {volume} {86}},\ \bibinfo {pages} {307}
  (\bibinfo {year} {2014})}\BibitemShut {NoStop}%
\bibitem [{\citenamefont {Kedem}\ and\ \citenamefont
  {Vaidman}(2010)}]{kedem_wv}%
  \BibitemOpen
  \bibfield  {author} {\bibinfo {author} {\bibfnamefont {Y.}~\bibnamefont
  {Kedem}}\ and\ \bibinfo {author} {\bibfnamefont {L.}~\bibnamefont
  {Vaidman}},\ }\href {\doibase 10.1103/PhysRevLett.105.230401} {\bibfield
  {journal} {\bibinfo  {journal} {Phys. Rev. Lett.}\ }\textbf {\bibinfo
  {volume} {105}},\ \bibinfo {pages} {230401} (\bibinfo {year}
  {2010})}\BibitemShut {NoStop}%
\bibitem [{\citenamefont {Dixon}\ \emph {et~al.}(2009)\citenamefont {Dixon},
  \citenamefont {Starling}, \citenamefont {Jordan},\ and\ \citenamefont
  {Howell}}]{ben_dixon}%
  \BibitemOpen
  \bibfield  {author} {\bibinfo {author} {\bibfnamefont {P.~B.}\ \bibnamefont
  {Dixon}}, \bibinfo {author} {\bibfnamefont {D.~J.}\ \bibnamefont {Starling}},
  \bibinfo {author} {\bibfnamefont {A.~N.}\ \bibnamefont {Jordan}}, \ and\
  \bibinfo {author} {\bibfnamefont {J.~C.}\ \bibnamefont {Howell}},\ }\href
  {\doibase 10.1103/PhysRevLett.102.173601} {\bibfield  {journal} {\bibinfo
  {journal} {Phys. Rev. Lett.}\ }\textbf {\bibinfo {volume} {102}},\ \bibinfo
  {pages} {173601} (\bibinfo {year} {2009})}\BibitemShut {NoStop}%
\bibitem [{\citenamefont {Salazar-Serrano}\ \emph {et~al.}(2015)\citenamefont
  {Salazar-Serrano}, \citenamefont {Barrera}, \citenamefont {Amaya},
  \citenamefont {Sales}, \citenamefont {Pruneri}, \citenamefont {Capmany},\
  and\ \citenamefont {Torres}}]{thermometry}%
  \BibitemOpen
  \bibfield  {author} {\bibinfo {author} {\bibfnamefont {L.}~\bibnamefont
  {Salazar-Serrano}}, \bibinfo {author} {\bibfnamefont {D.}~\bibnamefont
  {Barrera}}, \bibinfo {author} {\bibfnamefont {W.}~\bibnamefont {Amaya}},
  \bibinfo {author} {\bibfnamefont {S.}~\bibnamefont {Sales}}, \bibinfo
  {author} {\bibfnamefont {V.}~\bibnamefont {Pruneri}}, \bibinfo {author}
  {\bibfnamefont {J.}~\bibnamefont {Capmany}}, \ and\ \bibinfo {author}
  {\bibfnamefont {J.}~\bibnamefont {Torres}},\ }\href
  {https://www.osapublishing.org/ol/abstract.cfm?uri=ol-40-17-3962} {\bibfield
  {journal} {\bibinfo  {journal} {Optics letters}\ }\textbf {\bibinfo {volume}
  {40}},\ \bibinfo {pages} {3962} (\bibinfo {year} {2015})}\BibitemShut
  {NoStop}%
\bibitem [{\citenamefont {Pati}\ \emph {et~al.}(2020)\citenamefont {Pati},
  \citenamefont {Mukhopadhyay}, \citenamefont {Chakraborty},\ and\
  \citenamefont {Ghosh}}]{sibasish}%
  \BibitemOpen
  \bibfield  {author} {\bibinfo {author} {\bibfnamefont {A.~K.}\ \bibnamefont
  {Pati}}, \bibinfo {author} {\bibfnamefont {C.}~\bibnamefont {Mukhopadhyay}},
  \bibinfo {author} {\bibfnamefont {S.}~\bibnamefont {Chakraborty}}, \ and\
  \bibinfo {author} {\bibfnamefont {S.}~\bibnamefont {Ghosh}},\ }\href
  {\doibase 10.1103/PhysRevA.102.012204} {\bibfield  {journal} {\bibinfo
  {journal} {Phys. Rev. A}\ }\textbf {\bibinfo {volume} {102}},\ \bibinfo
  {pages} {012204} (\bibinfo {year} {2020})}\BibitemShut {NoStop}%
\bibitem [{\citenamefont {Ghosh}\ \emph {et~al.}(2019)\citenamefont {Ghosh},
  \citenamefont {Kwek}, \citenamefont {Terno},\ and\ \citenamefont
  {Vinjanampathy}}]{ghosh2019weak}%
  \BibitemOpen
  \bibfield  {author} {\bibinfo {author} {\bibfnamefont {S.}~\bibnamefont
  {Ghosh}}, \bibinfo {author} {\bibfnamefont {L.-C.}\ \bibnamefont {Kwek}},
  \bibinfo {author} {\bibfnamefont {D.~R.}\ \bibnamefont {Terno}}, \ and\
  \bibinfo {author} {\bibfnamefont {S.}~\bibnamefont {Vinjanampathy}},\ }\href
  {https://arxiv.org/pdf/1912.10693.pdf} {\bibfield  {journal} {\bibinfo
  {journal} {arXiv preprint arXiv:1912.10693}\ } (\bibinfo {year}
  {2019})}\BibitemShut {NoStop}%
\bibitem [{\citenamefont {Pang}\ \emph {et~al.}(2014)\citenamefont {Pang},
  \citenamefont {Dressel},\ and\ \citenamefont {Brun}}]{pang2014entanglement}%
  \BibitemOpen
  \bibfield  {author} {\bibinfo {author} {\bibfnamefont {S.}~\bibnamefont
  {Pang}}, \bibinfo {author} {\bibfnamefont {J.}~\bibnamefont {Dressel}}, \
  and\ \bibinfo {author} {\bibfnamefont {T.~A.}\ \bibnamefont {Brun}},\ }\href
  {\doibase 10.1103/PhysRevLett.113.030401} {\bibfield  {journal} {\bibinfo
  {journal} {Phys. Rev. Lett.}\ }\textbf {\bibinfo {volume} {113}},\ \bibinfo
  {pages} {030401} (\bibinfo {year} {2014})}\BibitemShut {NoStop}%
\bibitem [{\citenamefont {Brun}\ \emph {et~al.}(2008)\citenamefont {Brun},
  \citenamefont {Di\'osi},\ and\ \citenamefont {Strunz}}]{PhysRevA.77.032101}%
  \BibitemOpen
  \bibfield  {author} {\bibinfo {author} {\bibfnamefont {T.~A.}\ \bibnamefont
  {Brun}}, \bibinfo {author} {\bibfnamefont {L.}~\bibnamefont {Di\'osi}}, \
  and\ \bibinfo {author} {\bibfnamefont {W.~T.}\ \bibnamefont {Strunz}},\
  }\href {\doibase 10.1103/PhysRevA.77.032101} {\bibfield  {journal} {\bibinfo
  {journal} {Phys. Rev. A}\ }\textbf {\bibinfo {volume} {77}},\ \bibinfo
  {pages} {032101} (\bibinfo {year} {2008})}\BibitemShut {NoStop}%
\bibitem [{\citenamefont {Boixo}\ \emph {et~al.}(2008)\citenamefont {Boixo},
  \citenamefont {Datta}, \citenamefont {Davis}, \citenamefont {Flammia},
  \citenamefont {Shaji},\ and\ \citenamefont {Caves}}]{Caves}%
  \BibitemOpen
  \bibfield  {author} {\bibinfo {author} {\bibfnamefont {S.}~\bibnamefont
  {Boixo}}, \bibinfo {author} {\bibfnamefont {A.}~\bibnamefont {Datta}},
  \bibinfo {author} {\bibfnamefont {M.~J.}\ \bibnamefont {Davis}}, \bibinfo
  {author} {\bibfnamefont {S.~T.}\ \bibnamefont {Flammia}}, \bibinfo {author}
  {\bibfnamefont {A.}~\bibnamefont {Shaji}}, \ and\ \bibinfo {author}
  {\bibfnamefont {C.~M.}\ \bibnamefont {Caves}},\ }\href {\doibase
  10.1103/PhysRevLett.101.040403} {\bibfield  {journal} {\bibinfo  {journal}
  {Phys. Rev. Lett.}\ }\textbf {\bibinfo {volume} {101}},\ \bibinfo {pages}
  {040403} (\bibinfo {year} {2008})}\BibitemShut {NoStop}%
\bibitem [{\citenamefont {Zwierz}\ \emph {et~al.}(2010)\citenamefont {Zwierz},
  \citenamefont {P\'erez-Delgado},\ and\ \citenamefont
  {Kok}}]{zwierz2010general}%
  \BibitemOpen
  \bibfield  {author} {\bibinfo {author} {\bibfnamefont {M.}~\bibnamefont
  {Zwierz}}, \bibinfo {author} {\bibfnamefont {C.~A.}\ \bibnamefont
  {P\'erez-Delgado}}, \ and\ \bibinfo {author} {\bibfnamefont {P.}~\bibnamefont
  {Kok}},\ }\href {\doibase 10.1103/PhysRevLett.105.180402} {\bibfield
  {journal} {\bibinfo  {journal} {Phys. Rev. Lett.}\ }\textbf {\bibinfo
  {volume} {105}},\ \bibinfo {pages} {180402} (\bibinfo {year}
  {2010})}\BibitemShut {NoStop}%
\bibitem [{\citenamefont {Binder}\ \emph {et~al.}(2015)\citenamefont {Binder},
  \citenamefont {Vinjanampathy}, \citenamefont {Modi},\ and\ \citenamefont
  {Goold}}]{binder2015quantacell}%
  \BibitemOpen
  \bibfield  {author} {\bibinfo {author} {\bibfnamefont {F.~C.}\ \bibnamefont
  {Binder}}, \bibinfo {author} {\bibfnamefont {S.}~\bibnamefont
  {Vinjanampathy}}, \bibinfo {author} {\bibfnamefont {K.}~\bibnamefont {Modi}},
  \ and\ \bibinfo {author} {\bibfnamefont {J.}~\bibnamefont {Goold}},\ }\href
  {\doibase 10.1088/1367-2630/17/7/075015} {\bibfield  {journal} {\bibinfo
  {journal} {New Journal of Physics}\ }\textbf {\bibinfo {volume} {17}},\
  \bibinfo {pages} {075015} (\bibinfo {year} {2015})}\BibitemShut {NoStop}%
\bibitem [{\citenamefont {Campaioli}\ \emph {et~al.}(2017)\citenamefont
  {Campaioli}, \citenamefont {Pollock}, \citenamefont {Binder}, \citenamefont
  {C\'eleri}, \citenamefont {Goold}, \citenamefont {Vinjanampathy},\ and\
  \citenamefont {Modi}}]{campaioli2017enhancing}%
  \BibitemOpen
  \bibfield  {author} {\bibinfo {author} {\bibfnamefont {F.}~\bibnamefont
  {Campaioli}}, \bibinfo {author} {\bibfnamefont {F.~A.}\ \bibnamefont
  {Pollock}}, \bibinfo {author} {\bibfnamefont {F.~C.}\ \bibnamefont {Binder}},
  \bibinfo {author} {\bibfnamefont {L.}~\bibnamefont {C\'eleri}}, \bibinfo
  {author} {\bibfnamefont {J.}~\bibnamefont {Goold}}, \bibinfo {author}
  {\bibfnamefont {S.}~\bibnamefont {Vinjanampathy}}, \ and\ \bibinfo {author}
  {\bibfnamefont {K.}~\bibnamefont {Modi}},\ }\href {\doibase
  10.1103/PhysRevLett.118.150601} {\bibfield  {journal} {\bibinfo  {journal}
  {Phys. Rev. Lett.}\ }\textbf {\bibinfo {volume} {118}},\ \bibinfo {pages}
  {150601} (\bibinfo {year} {2017})}\BibitemShut {NoStop}%
\bibitem [{\citenamefont {Rossini}\ \emph {et~al.}(2019)\citenamefont
  {Rossini}, \citenamefont {Andolina},\ and\ \citenamefont
  {Polini}}]{rossini2019many}%
  \BibitemOpen
  \bibfield  {author} {\bibinfo {author} {\bibfnamefont {D.}~\bibnamefont
  {Rossini}}, \bibinfo {author} {\bibfnamefont {G.~M.}\ \bibnamefont
  {Andolina}}, \ and\ \bibinfo {author} {\bibfnamefont {M.}~\bibnamefont
  {Polini}},\ }\href {\doibase 10.1103/PhysRevB.100.115142} {\bibfield
  {journal} {\bibinfo  {journal} {Phys. Rev. B}\ }\textbf {\bibinfo {volume}
  {100}},\ \bibinfo {pages} {115142} (\bibinfo {year} {2019})}\BibitemShut
  {NoStop}%
\bibitem [{\citenamefont {Andolina}\ \emph {et~al.}(2019)\citenamefont
  {Andolina}, \citenamefont {Keck}, \citenamefont {Mari}, \citenamefont
  {Campisi}, \citenamefont {Giovannetti},\ and\ \citenamefont
  {Polini}}]{PhysRevLett.122.047702}%
  \BibitemOpen
  \bibfield  {author} {\bibinfo {author} {\bibfnamefont {G.~M.}\ \bibnamefont
  {Andolina}}, \bibinfo {author} {\bibfnamefont {M.}~\bibnamefont {Keck}},
  \bibinfo {author} {\bibfnamefont {A.}~\bibnamefont {Mari}}, \bibinfo {author}
  {\bibfnamefont {M.}~\bibnamefont {Campisi}}, \bibinfo {author} {\bibfnamefont
  {V.}~\bibnamefont {Giovannetti}}, \ and\ \bibinfo {author} {\bibfnamefont
  {M.}~\bibnamefont {Polini}},\ }\href {\doibase
  10.1103/PhysRevLett.122.047702} {\bibfield  {journal} {\bibinfo  {journal}
  {Phys. Rev. Lett.}\ }\textbf {\bibinfo {volume} {122}},\ \bibinfo {pages}
  {047702} (\bibinfo {year} {2019})}\BibitemShut {NoStop}%
\bibitem [{\citenamefont {Giovannetti}\ \emph {et~al.}(2004)\citenamefont
  {Giovannetti}, \citenamefont {Lloyd},\ and\ \citenamefont
  {Maccone}}]{Giovannetti1330}%
  \BibitemOpen
  \bibfield  {author} {\bibinfo {author} {\bibfnamefont {V.}~\bibnamefont
  {Giovannetti}}, \bibinfo {author} {\bibfnamefont {S.}~\bibnamefont {Lloyd}},
  \ and\ \bibinfo {author} {\bibfnamefont {L.}~\bibnamefont {Maccone}},\ }\href
  {\doibase 10.1126/science.1104149} {\bibfield  {journal} {\bibinfo  {journal}
  {Science}\ }\textbf {\bibinfo {volume} {306}},\ \bibinfo {pages} {1330}
  (\bibinfo {year} {2004})}\BibitemShut {NoStop}%
\bibitem [{\citenamefont {B{\"a}rtschi}\ and\ \citenamefont
  {Eidenbenz}(2019)}]{bartschi2019deterministic}%
  \BibitemOpen
  \bibfield  {author} {\bibinfo {author} {\bibfnamefont {A.}~\bibnamefont
  {B{\"a}rtschi}}\ and\ \bibinfo {author} {\bibfnamefont {S.}~\bibnamefont
  {Eidenbenz}},\ }in\ \href@noop {} {\emph {\bibinfo {booktitle} {Fundamentals
  of Computation Theory}}},\ \bibinfo {editor} {edited by\ \bibinfo {editor}
  {\bibfnamefont {L.~A.}\ \bibnamefont {Gasieniec}}, \bibinfo {editor}
  {\bibfnamefont {J.}~\bibnamefont {Jansson}}, \ and\ \bibinfo {editor}
  {\bibfnamefont {C.}~\bibnamefont {Levcopoulos}}}\ (\bibinfo  {publisher}
  {Springer International Publishing},\ \bibinfo {year} {2019})\ pp.\ \bibinfo
  {pages} {126--139}\BibitemShut {NoStop}%
\bibitem [{\citenamefont {Garbe}\ \emph {et~al.}(2017)\citenamefont {Garbe},
  \citenamefont {Egusquiza}, \citenamefont {Solano}, \citenamefont {Ciuti},
  \citenamefont {Coudreau}, \citenamefont {Milman},\ and\ \citenamefont
  {Felicetti}}]{Solano_two_photon_dicke_model}%
  \BibitemOpen
  \bibfield  {author} {\bibinfo {author} {\bibfnamefont {L.}~\bibnamefont
  {Garbe}}, \bibinfo {author} {\bibfnamefont {I.~L.}\ \bibnamefont
  {Egusquiza}}, \bibinfo {author} {\bibfnamefont {E.}~\bibnamefont {Solano}},
  \bibinfo {author} {\bibfnamefont {C.}~\bibnamefont {Ciuti}}, \bibinfo
  {author} {\bibfnamefont {T.}~\bibnamefont {Coudreau}}, \bibinfo {author}
  {\bibfnamefont {P.}~\bibnamefont {Milman}}, \ and\ \bibinfo {author}
  {\bibfnamefont {S.}~\bibnamefont {Felicetti}},\ }\href {\doibase
  10.1103/PhysRevA.95.053854} {\bibfield  {journal} {\bibinfo  {journal} {Phys.
  Rev. A}\ }\textbf {\bibinfo {volume} {95}},\ \bibinfo {pages} {053854}
  (\bibinfo {year} {2017})}\BibitemShut {NoStop}%
\bibitem [{\citenamefont {Garbe}\ \emph {et~al.}(2020)\citenamefont {Garbe},
  \citenamefont {Wade}, \citenamefont {Minganti}, \citenamefont {Shammah},
  \citenamefont {Felicetti},\ and\ \citenamefont {Nori}}]{Garbe_2020}%
  \BibitemOpen
  \bibfield  {author} {\bibinfo {author} {\bibfnamefont {L.}~\bibnamefont
  {Garbe}}, \bibinfo {author} {\bibfnamefont {P.}~\bibnamefont {Wade}},
  \bibinfo {author} {\bibfnamefont {F.}~\bibnamefont {Minganti}}, \bibinfo
  {author} {\bibfnamefont {N.}~\bibnamefont {Shammah}}, \bibinfo {author}
  {\bibfnamefont {S.}~\bibnamefont {Felicetti}}, \ and\ \bibinfo {author}
  {\bibfnamefont {F.}~\bibnamefont {Nori}},\ }\href {\doibase
  10.1038/s41598-020-69704-6} {\bibfield  {journal} {\bibinfo  {journal}
  {Scientific Reports}\ }\textbf {\bibinfo {volume} {10}} (\bibinfo {year}
  {2020}),\ 10.1038/s41598-020-69704-6}\BibitemShut {NoStop}%
\bibitem [{\citenamefont {Fink}\ \emph {et~al.}(2009)\citenamefont {Fink},
  \citenamefont {Bianchetti}, \citenamefont {Baur}, \citenamefont {G\"oppl},
  \citenamefont {Steffen}, \citenamefont {Filipp}, \citenamefont {Leek},
  \citenamefont {Blais},\ and\ \citenamefont {Wallraff}}]{fink2009dressed}%
  \BibitemOpen
  \bibfield  {author} {\bibinfo {author} {\bibfnamefont {J.~M.}\ \bibnamefont
  {Fink}}, \bibinfo {author} {\bibfnamefont {R.}~\bibnamefont {Bianchetti}},
  \bibinfo {author} {\bibfnamefont {M.}~\bibnamefont {Baur}}, \bibinfo {author}
  {\bibfnamefont {M.}~\bibnamefont {G\"oppl}}, \bibinfo {author} {\bibfnamefont
  {L.}~\bibnamefont {Steffen}}, \bibinfo {author} {\bibfnamefont
  {S.}~\bibnamefont {Filipp}}, \bibinfo {author} {\bibfnamefont {P.~J.}\
  \bibnamefont {Leek}}, \bibinfo {author} {\bibfnamefont {A.}~\bibnamefont
  {Blais}}, \ and\ \bibinfo {author} {\bibfnamefont {A.}~\bibnamefont
  {Wallraff}},\ }\href {\doibase 10.1103/PhysRevLett.103.083601} {\bibfield
  {journal} {\bibinfo  {journal} {Phys. Rev. Lett.}\ }\textbf {\bibinfo
  {volume} {103}},\ \bibinfo {pages} {083601} (\bibinfo {year}
  {2009})}\BibitemShut {NoStop}%
\bibitem [{\citenamefont {Mlynek}\ \emph {et~al.}(2014)\citenamefont {Mlynek},
  \citenamefont {Abdumalikov}, \citenamefont {Eichler},\ and\ \citenamefont
  {Wallraff}}]{mlynek2014observation}%
  \BibitemOpen
  \bibfield  {author} {\bibinfo {author} {\bibfnamefont {J.~A.}\ \bibnamefont
  {Mlynek}}, \bibinfo {author} {\bibfnamefont {A.~A.}\ \bibnamefont
  {Abdumalikov}}, \bibinfo {author} {\bibfnamefont {C.}~\bibnamefont
  {Eichler}}, \ and\ \bibinfo {author} {\bibfnamefont {A.}~\bibnamefont
  {Wallraff}},\ }\href {\doibase 10.1038/ncomms6186} {\bibfield  {journal}
  {\bibinfo  {journal} {Nature Communications}\ }\textbf {\bibinfo {volume}
  {5}},\ \bibinfo {pages} {5186} (\bibinfo {year} {2014})}\BibitemShut
  {NoStop}%
\bibitem [{\citenamefont {Song}\ \emph {et~al.}(2019)\citenamefont {Song},
  \citenamefont {Xu}, \citenamefont {Li}, \citenamefont {Zhang}, \citenamefont
  {Zhang}, \citenamefont {Liu}, \citenamefont {Guo}, \citenamefont {Wang},
  \citenamefont {Ren}, \citenamefont {Hao}, \citenamefont {Feng}, \citenamefont
  {Fan}, \citenamefont {Zheng}, \citenamefont {Wang}, \citenamefont {Wang},\
  and\ \citenamefont {Zhu}}]{Song574}%
  \BibitemOpen
  \bibfield  {author} {\bibinfo {author} {\bibfnamefont {C.}~\bibnamefont
  {Song}}, \bibinfo {author} {\bibfnamefont {K.}~\bibnamefont {Xu}}, \bibinfo
  {author} {\bibfnamefont {H.}~\bibnamefont {Li}}, \bibinfo {author}
  {\bibfnamefont {Y.-R.}\ \bibnamefont {Zhang}}, \bibinfo {author}
  {\bibfnamefont {X.}~\bibnamefont {Zhang}}, \bibinfo {author} {\bibfnamefont
  {W.}~\bibnamefont {Liu}}, \bibinfo {author} {\bibfnamefont {Q.}~\bibnamefont
  {Guo}}, \bibinfo {author} {\bibfnamefont {Z.}~\bibnamefont {Wang}}, \bibinfo
  {author} {\bibfnamefont {W.}~\bibnamefont {Ren}}, \bibinfo {author}
  {\bibfnamefont {J.}~\bibnamefont {Hao}}, \bibinfo {author} {\bibfnamefont
  {H.}~\bibnamefont {Feng}}, \bibinfo {author} {\bibfnamefont {H.}~\bibnamefont
  {Fan}}, \bibinfo {author} {\bibfnamefont {D.}~\bibnamefont {Zheng}}, \bibinfo
  {author} {\bibfnamefont {D.-W.}\ \bibnamefont {Wang}}, \bibinfo {author}
  {\bibfnamefont {H.}~\bibnamefont {Wang}}, \ and\ \bibinfo {author}
  {\bibfnamefont {S.-Y.}\ \bibnamefont {Zhu}},\ }\href {\doibase
  10.1126/science.aay0600} {\bibfield  {journal} {\bibinfo  {journal}
  {Science}\ }\textbf {\bibinfo {volume} {365}},\ \bibinfo {pages} {574}
  (\bibinfo {year} {2019})}\BibitemShut {NoStop}%
\bibitem [{\citenamefont {Vijay}\ \emph {et~al.}(2011)\citenamefont {Vijay},
  \citenamefont {Slichter},\ and\ \citenamefont
  {Siddiqi}}]{vijay2011observation}%
  \BibitemOpen
  \bibfield  {author} {\bibinfo {author} {\bibfnamefont {R.}~\bibnamefont
  {Vijay}}, \bibinfo {author} {\bibfnamefont {D.~H.}\ \bibnamefont {Slichter}},
  \ and\ \bibinfo {author} {\bibfnamefont {I.}~\bibnamefont {Siddiqi}},\ }\href
  {\doibase 10.1103/PhysRevLett.106.110502} {\bibfield  {journal} {\bibinfo
  {journal} {Phys. Rev. Lett.}\ }\textbf {\bibinfo {volume} {106}},\ \bibinfo
  {pages} {110502} (\bibinfo {year} {2011})}\BibitemShut {NoStop}%
\bibitem [{\citenamefont {Pozar}(1990)}]{pozar1990microwave}%
  \BibitemOpen
  \bibfield  {author} {\bibinfo {author} {\bibfnamefont {D.}~\bibnamefont
  {Pozar}},\ }\href {https://books.google.co.in/books?id=0XLFQgAACAAJ} {\emph
  {\bibinfo {title} {Microwave Engineering}}},\ Addison-Wesley series in
  electrical and computer engineering\ (\bibinfo  {publisher}
  {Addison-Wesley},\ \bibinfo {year} {1990})\BibitemShut {NoStop}%
\bibitem [{\citenamefont {Felicetti}\ \emph {et~al.}(2018)\citenamefont
  {Felicetti}, \citenamefont {Rossatto}, \citenamefont {Rico}, \citenamefont
  {Solano},\ and\ \citenamefont
  {Forn-D\'{\i}az}}]{Solano_two_photon_rabi_model}%
  \BibitemOpen
  \bibfield  {author} {\bibinfo {author} {\bibfnamefont {S.}~\bibnamefont
  {Felicetti}}, \bibinfo {author} {\bibfnamefont {D.~Z.}\ \bibnamefont
  {Rossatto}}, \bibinfo {author} {\bibfnamefont {E.}~\bibnamefont {Rico}},
  \bibinfo {author} {\bibfnamefont {E.}~\bibnamefont {Solano}}, \ and\ \bibinfo
  {author} {\bibfnamefont {P.}~\bibnamefont {Forn-D\'{\i}az}},\ }\href
  {\doibase 10.1103/PhysRevA.97.013851} {\bibfield  {journal} {\bibinfo
  {journal} {Phys. Rev. A}\ }\textbf {\bibinfo {volume} {97}},\ \bibinfo
  {pages} {013851} (\bibinfo {year} {2018})}\BibitemShut {NoStop}%
\bibitem [{\citenamefont {James}\ and\ \citenamefont
  {Jerke}(2007)}]{james2007effective}%
  \BibitemOpen
  \bibfield  {author} {\bibinfo {author} {\bibfnamefont {D.~F.}\ \bibnamefont
  {James}}\ and\ \bibinfo {author} {\bibfnamefont {J.}~\bibnamefont {Jerke}},\
  }\href {\doibase 10.1139/p07-060} {\bibfield  {journal} {\bibinfo  {journal}
  {Canadian Journal of Physics}\ }\textbf {\bibinfo {volume} {85}},\ \bibinfo
  {pages} {625} (\bibinfo {year} {2007})}\BibitemShut {NoStop}%
\bibitem [{\citenamefont {Shao}\ \emph {et~al.}(2017)\citenamefont {Shao},
  \citenamefont {Wu},\ and\ \citenamefont {Feng}}]{PhysRevA.95.032124}%
  \BibitemOpen
  \bibfield  {author} {\bibinfo {author} {\bibfnamefont {W.}~\bibnamefont
  {Shao}}, \bibinfo {author} {\bibfnamefont {C.}~\bibnamefont {Wu}}, \ and\
  \bibinfo {author} {\bibfnamefont {X.-L.}\ \bibnamefont {Feng}},\ }\href
  {\doibase 10.1103/PhysRevA.95.032124} {\bibfield  {journal} {\bibinfo
  {journal} {Phys. Rev. A}\ }\textbf {\bibinfo {volume} {95}},\ \bibinfo
  {pages} {032124} (\bibinfo {year} {2017})}\BibitemShut {NoStop}%
\bibitem [{\citenamefont {Forn-D\'{\i}az}\ \emph {et~al.}(2010)\citenamefont
  {Forn-D\'{\i}az}, \citenamefont {Lisenfeld}, \citenamefont {Marcos},
  \citenamefont {Garc\'{\i}a-Ripoll}, \citenamefont {Solano}, \citenamefont
  {Harmans},\ and\ \citenamefont {Mooij}}]{forn2010observation}%
  \BibitemOpen
  \bibfield  {author} {\bibinfo {author} {\bibfnamefont {P.}~\bibnamefont
  {Forn-D\'{\i}az}}, \bibinfo {author} {\bibfnamefont {J.}~\bibnamefont
  {Lisenfeld}}, \bibinfo {author} {\bibfnamefont {D.}~\bibnamefont {Marcos}},
  \bibinfo {author} {\bibfnamefont {J.~J.}\ \bibnamefont {Garc\'{\i}a-Ripoll}},
  \bibinfo {author} {\bibfnamefont {E.}~\bibnamefont {Solano}}, \bibinfo
  {author} {\bibfnamefont {C.~J. P.~M.}\ \bibnamefont {Harmans}}, \ and\
  \bibinfo {author} {\bibfnamefont {J.~E.}\ \bibnamefont {Mooij}},\ }\href
  {\doibase 10.1103/PhysRevLett.105.237001} {\bibfield  {journal} {\bibinfo
  {journal} {Phys. Rev. Lett.}\ }\textbf {\bibinfo {volume} {105}},\ \bibinfo
  {pages} {237001} (\bibinfo {year} {2010})}\BibitemShut {NoStop}%
\bibitem [{\citenamefont {Ma}\ \emph {et~al.}(2011)\citenamefont {Ma},
  \citenamefont {Wang}, \citenamefont {Sun},\ and\ \citenamefont
  {Nori}}]{ma2011quantum}%
  \BibitemOpen
  \bibfield  {author} {\bibinfo {author} {\bibfnamefont {J.}~\bibnamefont
  {Ma}}, \bibinfo {author} {\bibfnamefont {X.}~\bibnamefont {Wang}}, \bibinfo
  {author} {\bibfnamefont {C.}~\bibnamefont {Sun}}, \ and\ \bibinfo {author}
  {\bibfnamefont {F.}~\bibnamefont {Nori}},\ }\href {\doibase
  https://doi.org/10.1016/j.physrep.2011.08.003} {\bibfield  {journal}
  {\bibinfo  {journal} {Physics Reports}\ }\textbf {\bibinfo {volume} {509}},\
  \bibinfo {pages} {89 } (\bibinfo {year} {2011})}\BibitemShut {NoStop}%
\bibitem [{\citenamefont {Oszmaniec}\ \emph {et~al.}(2016)\citenamefont
  {Oszmaniec}, \citenamefont {Grudka}, \citenamefont {Horodecki},\ and\
  \citenamefont {Wójcik}}]{PhysRevLett.116.110403}%
  \BibitemOpen
  \bibfield  {author} {\bibinfo {author} {\bibfnamefont {M.}~\bibnamefont
  {Oszmaniec}}, \bibinfo {author} {\bibfnamefont {A.}~\bibnamefont {Grudka}},
  \bibinfo {author} {\bibfnamefont {M.}~\bibnamefont {Horodecki}}, \ and\
  \bibinfo {author} {\bibfnamefont {A.}~\bibnamefont {Wójcik}},\ }\href
  {\doibase 10.1103/PhysRevLett.116.110403} {\bibfield  {journal} {\bibinfo
  {journal} {Phys. Rev. Lett.}\ }\textbf {\bibinfo {volume} {116}},\ \bibinfo
  {pages} {110403} (\bibinfo {year} {2016})}\BibitemShut {NoStop}%
\bibitem [{\citenamefont {Li}\ \emph {et~al.}(2017)\citenamefont {Li},
  \citenamefont {Long}, \citenamefont {Katiyar}, \citenamefont {Xin},
  \citenamefont {Feng}, \citenamefont {Lu},\ and\ \citenamefont
  {Laflamme}}]{exptimplinitial}%
  \BibitemOpen
  \bibfield  {author} {\bibinfo {author} {\bibfnamefont {K.}~\bibnamefont
  {Li}}, \bibinfo {author} {\bibfnamefont {G.}~\bibnamefont {Long}}, \bibinfo
  {author} {\bibfnamefont {H.}~\bibnamefont {Katiyar}}, \bibinfo {author}
  {\bibfnamefont {T.}~\bibnamefont {Xin}}, \bibinfo {author} {\bibfnamefont
  {G.}~\bibnamefont {Feng}}, \bibinfo {author} {\bibfnamefont {D.}~\bibnamefont
  {Lu}}, \ and\ \bibinfo {author} {\bibfnamefont {R.}~\bibnamefont
  {Laflamme}},\ }\href {\doibase 10.1103/PhysRevA.95.022334} {\bibfield
  {journal} {\bibinfo  {journal} {Phys. Rev. A}\ }\textbf {\bibinfo {volume}
  {95}},\ \bibinfo {pages} {022334} (\bibinfo {year} {2017})}\BibitemShut
  {NoStop}%
\bibitem [{\citenamefont {Alves}\ \emph {et~al.}(2015)\citenamefont {Alves},
  \citenamefont {Escher}, \citenamefont {de~Matos~Filho}, \citenamefont
  {Zagury},\ and\ \citenamefont {Davidovich}}]{PhysRevA.91.062107}%
  \BibitemOpen
  \bibfield  {author} {\bibinfo {author} {\bibfnamefont {G.~B.}\ \bibnamefont
  {Alves}}, \bibinfo {author} {\bibfnamefont {B.~M.}\ \bibnamefont {Escher}},
  \bibinfo {author} {\bibfnamefont {R.~L.}\ \bibnamefont {de~Matos~Filho}},
  \bibinfo {author} {\bibfnamefont {N.}~\bibnamefont {Zagury}}, \ and\ \bibinfo
  {author} {\bibfnamefont {L.}~\bibnamefont {Davidovich}},\ }\href {\doibase
  10.1103/PhysRevA.91.062107} {\bibfield  {journal} {\bibinfo  {journal} {Phys.
  Rev. A}\ }\textbf {\bibinfo {volume} {91}},\ \bibinfo {pages} {062107}
  (\bibinfo {year} {2015})}\BibitemShut {NoStop}%
\bibitem [{\citenamefont {Braunstein}\ and\ \citenamefont
  {Caves}(1994)}]{PhysRevLett.72.3439}%
  \BibitemOpen
  \bibfield  {author} {\bibinfo {author} {\bibfnamefont {S.~L.}\ \bibnamefont
  {Braunstein}}\ and\ \bibinfo {author} {\bibfnamefont {C.~M.}\ \bibnamefont
  {Caves}},\ }\href {\doibase 10.1103/PhysRevLett.72.3439} {\bibfield
  {journal} {\bibinfo  {journal} {Phys. Rev. Lett.}\ }\textbf {\bibinfo
  {volume} {72}},\ \bibinfo {pages} {3439} (\bibinfo {year}
  {1994})}\BibitemShut {NoStop}%
\end{thebibliography}
	\color{RoyalBlue}

		\color{Black}
\clearpage
	\appendix
	\onecolumngrid
\section{A: Proof of Simultaneous Optimization}
The final state can be written as

\begin{equation}
\ket{\psi_{f}}=\sqrt{P_{s}^{(2j)}}\ket{\psi_{i}}+\sqrt{1-P_{s}^{(2j)}}\ket{\psi_{i}^{\perp}}, \label{a1}
\end{equation}

where $P_{s}^{(2j)}$ is small. In   Eq.~\eqref{a1} $\ket{\psi_{i}^{\perp}}$ is orthogonal to the state $\ket{\psi_{i}}$.  The orthogonal state can be written in terms of initial state $\ket{\psi_{i}}$ as

\begin{equation}
\ket{\psi_{i}^{\perp}} = \frac{({A}-\braket{A}_{\ket{\psi_{i}}})\ket{\psi_{i}}}{\sqrt{Var(A)_{\ket{\psi_{i}}}}}. \label{a2}
\end{equation} 

If we fix $P_{s}^{(2j)}=\kappa j^2$, the equation above can be rewritten as

\begin{equation}
\ket{\psi_{f}} = \sqrt{\kappa j^2} \frac{1}{\sqrt{2}} (\ket{j,0} + \ket{j,-j}) + \frac{\sqrt{1-\kappa j^2}}{\sqrt{Var(A)_{\ket{\psi_{i}}}}} \frac{1}{\sqrt{2}} \bigg[\frac{j^2}{2}\ket{j,0}-\frac{j^2}{2}\ket{j,-j}\bigg]. \label{a22}
\end{equation}
The variance of $A$ goes as $j^4/4$, so Eq.~\eqref{a22} becomes

\begin{equation}
    \ket{\psi_{f}} \propto \sqrt{\kappa}j (\ket{j,0}+\ket{j,-j}) + \sqrt{1-\kappa j^2} (\ket{j,0}-\ket{j,-j}).
\end{equation}
Since $P_{s}^{(2j)}$ is small, $\sqrt{1-\kappa j^2}$ is approximated to unity.  Finally we get the postselection state as

\begin{equation}
    \ket{\psi_{f}} \propto (\sqrt{\kappa}j+1)\ket{j,0} + (\sqrt{\kappa}j-1)\ket{j,-j}.  \label{a3}
\end{equation}

This is Eq.~(13) in the main text.

\section{B:  Preparation of Initial State}
We propose a non-deterministic scheme for preparing the initial state of the system.  The initial state can be constructed from the circuit diagram given in Fig.~(2). The Dicke states $\ket{j,m_1}$ and $\ket{j,m_2}$ are created by the action of unitaries $U_{j,m_1}$ and $U_{j,m_2}$ on the computational states $\ket{0}^{\otimes j-m_1} \otimes \ket{1}^{\otimes j+m_1}$ and $\ket{0}^{\otimes j-m_2} \otimes \ket{1}^{\otimes j+m_2}$.  Thus the input state of the circuit is
\begin{equation}
    \ket{R_1} = \ket{\nu} \otimes \ket{j,m_1} \otimes \ket{j,m_2}. \label{a4}
\end{equation}

The ancilla state $\ket{\nu}$ is taken to be $\alpha\ket{0}+\beta\ket{1}$. Setting $\alpha=\beta=1/\sqrt{2}$. The control-SWAP gate gets activated when state $\ket{1}$ passes through the first terminal.  The state after the control-SWAP gate is

\begin{equation}
    \ket{R_2}= \frac{1}{\sqrt{2}}(\ket{0} \otimes \ket{j,m_1} \otimes \ket{j,m_2}+\ket{1} \otimes \ket{j,m_2} \otimes \ket{j,m_1}). \label{a5}
\end{equation}

Finally the state is projected onto  $K=\ket{\mu}\bra{\mu}\otimes I\otimes\ket{\zeta}\bra{\zeta}$. For convenience, $\ket{\mu}$ is set to  $\vert\braket{j,m_1\vert \zeta}\vert\ket{0}+\vert\braket{j,m_2\vert \zeta}\vert\ket{1}$. The only condition required is the overlaps $\braket{j,m_1\vert\zeta}$ and $\braket{j,m_2\vert\zeta}$ are non-zero.  After measurement the state $\ket{R_{2}}$ becomes

\begin{equation}
    \ket{R_3} =  \ket{\mu} \otimes {\bigg[\frac{1}{\sqrt{2}} \frac{\braket{\zeta\vert j,m_2}}{\vert\braket{\zeta\vert j,m_2}\vert} \ket{j,m_1} + \frac{1}{\sqrt{2}} \frac{\braket{\zeta\vert j,m_1} }{\vert\braket{\zeta\vert j,m_1}\vert}\ket{j,m_2}\bigg]} \otimes \ket{\zeta}. \label{a66}
\end{equation}
For our problem, we choose $m_1 = 0$ and $m_2 = -j$ and the reference state $\ket{\zeta}$ as $\ket{+}^{\otimes 2j}$. The overlaps in Eq.~\eqref{a66} are $\braket{\zeta\vert j,0} = \vert\braket{\zeta\vert j,0}\vert = (1/\sqrt{2})^{2j} \binom{2j}{j}$ and $\braket{\zeta\vert j,-j} = \vert \braket{\zeta\vert j,-j}\vert = (1/\sqrt{2})^{2j}$.  Finally we get the input state from the output of second terminal as

\begin{equation}
\ket{\psi_{i}} = \frac{1}{\sqrt{2}}(\ket{j,0}+\ket{j,-j}). \label{a7}
\end{equation}

The success probability of this state preparation is evaluated to be
\begin{equation}
    P = \bra{R_{2}}K^{\dagger}K\ket{R_{2}} = \vert\braket{j,m_1\vert\zeta}\vert^2\vert\braket{j,m_2\vert\zeta}\vert^2. \label{a77}
\end{equation}
For the initial state defined in Eq.~(4), the success probability of state preparation is $P = (1/2)^{4j}\binom{2j}{j}^2$.
\section{C: Implementation of Measurement}

The post-measurement can be implemented using the circuit in Fig.~(3).  The input state of the circuit is
\begin{equation}
 \ket{R_1} = \ket{\mu} \otimes \ket{\Psi(t)} \otimes \ket{\zeta}. \nonumber
\end{equation}

The ancilla state is chosen to be $\ket{\mu} =\alpha\ket{0}+\beta\ket{1}$.  The control-SWAP gate gets activated when $\ket{1}$ passes through the first terminal.  The state after control-SWAP gate is

\begin{equation}
 \ket{R_2} = \alpha \ket{0} \otimes \ket{\Psi(t)} \otimes \ket{\zeta} + \beta \ket{1} \otimes \ket{\zeta} \otimes \ket{\Psi(t)}. \nonumber
\end{equation}

We finally measure the states with a joint quantum state $\ket{\nu} \otimes \ket{j,m_1} \otimes \ket{j,m_2}$ with the measurement operator defined as

\begin{equation}
   K = (\ket{\nu}\bra{\nu} \otimes \ket{j,m_1}\bra{j,m_1} \otimes \ket{j,m_2}\bra{j,m_2}). \label{a8}
\end{equation}

The state after measurement $K$ is
        \begin{equation}
       \ket{R_3} = \frac{(\ket{\nu}\bra{\nu} \otimes \ket{j,m_1}\bra{j,m_1} \otimes \ket{j,m_2}\bra{j,m_2})\ket{R_{2}}}{\sqrt{\bra{R_{2}}K^{\dagger}K\ket{R_{2}}}}, \nonumber
    \end{equation}
For convenience, we assume $\braket{\nu\vert 0} = \lambda  \braket{j,m_1\vert\zeta}$ and $\braket{\nu\vert 1} = \lambda \braket{j,m_2\vert\zeta}$, we get

\begin{equation}
 \ket{R_3} = \frac{\lambda \braket{j,m_1\vert\zeta} \braket{j,m_2\vert\zeta} [\alpha\bra{j,m_1}+\beta\bra{j,m_2}]\ket{\Psi(t)}  \ket{\nu} \otimes \ket{j,m_1} \otimes \ket{j,m_2}}{\sqrt{\bra{R_{2}}K^{\dagger}K\ket{R_{2}}}}. \label{a9}
\end{equation}

$\lambda$ can be obtained by normalizing $\ket{\nu}$.  On normalizing $\ket{\nu}$, we get
\begin{equation}
    \lambda =\frac{1}{\sqrt{\vert\braket{\zeta\vert j,m_1}\vert^2+\vert\braket{\zeta\vert j,m_2}\vert^2}}.
\end{equation}

The probability of the measurement  Eq.~\eqref{a8} for a particular outcome can be obtained from evaluating $\tilde{P}_{s}=\bra{R_{2}}K^{\dagger}K\ket{R_{2}}$ as follows

\begin{eqnarray}
    \tilde{P}_{s} & = & \bra{R_{2}}K^{\dagger}K\ket{R_{2}}, \nonumber \\
    & = & \bigg(\bra{j,m_2}\otimes\bra{j,m_1}\otimes\bra{\nu}\bigg[\bra{\Psi(t)}[\alpha^{*}\ket{j,m_1}+\beta^{*} \ket{j,m_2}]\frac{\braket{\zeta \vert j,m_1}\braket{\zeta\vert j,m_2}}{\sqrt{\vert\braket{\zeta\vert j,m_1}\vert^2+\vert\braket{\zeta\vert j,m_2}\vert^2}}\bigg]\bigg) \nonumber \\
    & &\times \bigg(\bigg[\frac{ \braket{j,m_1\vert\zeta} \braket{j,m_2\vert\zeta} }{\sqrt{\vert\braket{\zeta\vert j,m_1}\vert^2+\vert\braket{\zeta\vert j,m_2}\vert^2}} [\alpha\bra{j,m_1}+\beta\bra{j,m_2}]\ket{\Psi(t)}\bigg]   \ket{\nu} \otimes \ket{j,m_1} \otimes \ket{j,m_2} \bigg), \nonumber \\
    \tilde{P}_{s}& = & \bigg[\frac{ \vert\braket{j,m_1
    \vert\zeta}\vert^2 \vert\braket{j,m_2\vert\zeta}\vert^2 }{\vert\braket{\zeta\vert j,m_1}\vert^2+\vert\braket{\zeta\vert j,m_2}\vert^2}\bigg]
    [\alpha^{*}\braket{\Psi(t) \vert j,m_1}+\beta^{*} \braket{\Psi(t)\vert j,m_2}][\alpha\braket{j,m_1\vert\Psi(t)}+\beta\braket{j,m_2\vert\Psi(t)}]. \label{a10}
\end{eqnarray}

For postselecting onto the state given in Eq.~(13), we set $\alpha = (\sqrt{\kappa}j+1)/\sqrt{2(1+\kappa j^2)}$ and $\beta = (\sqrt{\kappa}j-1)/\sqrt{2(1+\kappa j^2)}$, $m_1 = 0$ and $m_2 = -j$. Choosing reference state as the equal superposition of Dicke states (i.e, $\ket{\zeta} = 1/\sqrt{2}[\ket{j,0}+\ket{j,-j}]$), which can be prepared with the circuit in the Fig.~(2) with low probability by choosing the reference state as $\ket{+}^{\otimes 2j}$ and using it as an offline resource.  Therefore the overlaps are $\vert\braket{j,m_1\vert\zeta}\vert=\vert\braket{j,m_2\vert\zeta}\vert=1/\sqrt{2}$.  The input state of the circuit is $\ket{\Psi(t)} = \exp(-ig A\otimes B)\ket{\psi_{i}}\otimes\ket{\phi_{i}} \approx(1 - igA \otimes B)\ket{\psi_{i}}\otimes\ket{\phi_{i}}$ for small values of $g$. Before evaluating the probability, we first calculate the following overlaps

\begin{eqnarray}
    \alpha^{*}\braket{\Psi(t)\vert j,m_1}+\beta^{*}\braket{\Psi(t)\vert j,m_2} & \approx & \bra{\phi_{i}} \bigg(\frac{1}{\sqrt{1+\kappa j^2}} (\sqrt{\kappa}j+i\frac{gB}{2}[j(j+1)(\sqrt{
    \kappa}j+1)+j(\sqrt{\kappa}j-1)])\bigg), \nonumber \\
    &\approx& \bra{\phi_{i}} \bigg(\frac{\sqrt{\kappa}j}{\sqrt{1+\kappa j^2}} (1+i\frac{g B}{2}[j^2+2j+j/\sqrt{\kappa}])\bigg), \nonumber \\
    &=& \bra{\phi_{i}}\bigg(\frac{\sqrt{\kappa}j}{\sqrt{1+\kappa j^2}}\exp\bigg(i\frac{g}{2}\bigg[j^2+2j+\frac{j}{\sqrt{\kappa}}\bigg]B\bigg)\bigg). \label{a100}
\end{eqnarray}
Since the assumption $\kappa j^2$ is very small, $j/\sqrt{\kappa}$ is the dominant term in the exponential.  So the overlap can be written as
\begin{equation}
    \alpha^{*}\braket{\Psi(t)\vert j,m_1}+\beta^{*}\braket{\Psi(t)\vert j,m_2} \approx \bra{\phi_{i}}\bigg(\frac{\sqrt{\kappa }j}{\sqrt{1+\kappa j^2}} \exp(ig\mathcal{A}_{w}B)\bigg). \label{a1001}
\end{equation}
Similarly the other overlap can also be evaluated in the same fashion 
\begin{eqnarray}
    \alpha\braket{j,m_1\vert\Psi(t)}+\beta\braket{j,m_2\vert\Psi(t)}
    & \approx &  \bigg(\frac{1}{\sqrt{1+\kappa j^2}} ( \sqrt{\kappa}j-i\frac{gB}{2}[j(j+1)(\sqrt{\kappa}j+1)+j(\sqrt{\kappa}j-1)])\bigg)\ket{\phi_{i}}, \nonumber \\
    &\approx&  \bigg(\frac{\sqrt{\kappa}j}{\sqrt{1+\kappa j^2}} (1-i\frac{g B}{2}[j^2+2j+j/\sqrt{\kappa}])\bigg)\ket{\phi_{i}}, \nonumber \\
    &=& \bigg(\frac{\sqrt{\kappa}j}{\sqrt{1+\kappa j^2}}\exp\bigg(-i\frac{g}{2}\bigg[j^2+2j+\frac{j}{\sqrt{\kappa}}\bigg]B\bigg)\bigg)\ket{\phi_{i}}, \nonumber \\
     \alpha\braket{j,m_1\vert\Psi(t)}+\beta\braket{j,m_2\vert\Psi(t)}
    & \approx & \bigg(\frac{\sqrt{\kappa}j}{\sqrt{1+\kappa j^2}} \exp(-ig\mathcal{A}_{w}B)\bigg)\ket{\phi_{i}}.
    \label{a101}
\end{eqnarray}
Substituting Eq.~\eqref{a1001} and Eq.~\eqref{a101} in Eq.~\eqref{a10}, we get

\begin{equation}
    \tilde{P}_{s} \propto  \frac{\kappa j^2}{4(1+\kappa j^2)}.
\end{equation}

Approximating $1+\kappa j^2$ to unity, the probability of measurement is $
\tilde{P}_{s} \propto  \kappa j^2$, which has the same quadratic scaling in $j$ as $P_{s}^{(2j)}$.

\end{document}